\documentclass[12pt]{article}
\pdfoutput=1

\usepackage{amsmath}
\usepackage{amssymb}
\usepackage{epsfig}
\usepackage{graphicx}
\usepackage{cite}
\usepackage{multirow}
\usepackage{longtable}
\usepackage{lscape}
\usepackage{bbm}
\usepackage{dcolumn}
\usepackage{rotating}

\allowdisplaybreaks[4] 

\newcommand{\capdef}{}
\newcommand{\mycaption}[2][\capdef]{\renewcommand{\capdef}{#2}%
        \caption[#1]{{\footnotesize #2}}}
\makeatletter
\renewcommand{\fnum@table}{\textbf{\tablename~\thetable}}
\renewcommand{\fnum@figure}{\textbf{\figurename~\thefigure}}
\makeatother

\newcounter{myenumi}

\renewcommand{\themyenumi}{\roman{myenumi}}
{\end{list}}

\newlength{\myem}
\newcounter{mysubequation}[equation]

\makeatletter
\renewcommand{\section}{\@startsection{section}{1}{0em}{-\baselineskip}%
{\baselineskip}{\normalfont\large\bfseries}}
\renewcommand{\subsection}%
{\@startsection{subsection}{2}{0em}{-0.7\baselineskip}%
{0.7\baselineskip}{\normalfont\bfseries}}
\makeatother

\newcommand{\bi}{\begin{itemize}}
\newcommand{\ei}{\end{itemize}}

\newcommand{\be}{\begin{equation}}
\newcommand{\ee}{\end{equation}}
\newcommand{\bea}{\begin{eqnarray}}
\newcommand{\eea}{\end{eqnarray}}

\newcommand{\stheta}{\sin^2 2 \theta_{13}}

\newcommand{\ie}{{\it i.e.}}

\newcommand{\cf}{{\it cf.}}

\newcommand{\eq}{Eq.}

\newcommand{\fig}{Fig.}

\newcommand{\Ref}{Ref.}
\newcommand{\Refs}{Refs.}

\newcommand{\Tab}{Table}

\newcommand{\equ}[1]{\eq~(\ref{equ:#1})}
\newcommand{\figu}[1]{\fig~\ref{fig:#1}}

\begin{document}

\begin{titlepage}

\renewcommand{\thefootnote}{\alph{footnote}}

\vspace*{-3.cm}
\begin{flushright}
SISSA  83/2010/EP
\end{flushright}

\vspace*{0.5cm}

\renewcommand{\thefootnote}{\fnsymbol{footnote}}
\setcounter{footnote}{-1}

{\begin{center}
{\large\bf
Perturbing exactly tri-bimaximal neutrino mixings\\ with charged lepton mass matrices
}

\end{center}}

\renewcommand{\thefootnote}{\alph{footnote}}

\vspace*{.8cm}
\vspace*{.3cm}
{\begin{center} {\large{\sc
                Davide~Meloni\footnote[1]{\makebox[1.cm]{Email:}
                davide.meloni@physik.uni-wuerzburg.de},
Florian~Plentinger\footnote[2]{\makebox[1.cm]{Email:}
                plentinger@sissa.it}, and
                Walter~Winter\footnote[3]{\makebox[1.cm]{Email:}
                winter@physik.uni-wuerzburg.de}
                }}
\end{center}}
\vspace*{0cm}
{\it
\begin{center}

\footnotemark[1]${}^,$\footnotemark[3]
       Institut f{\"u}r Theoretische Physik und Astrophysik, \\ Universit{\"a}t W{\"u}rzburg,
       97074 W{\"u}rzburg, Germany

\footnotemark[2]
      {SISSA and INFN-Sezione di Trieste},\\
{via Bonomea 265, 34136 Trieste, Italy}\\
\end{center}}

\vspace*{1.5cm}


{\Large \bf
\begin{center} Abstract \end{center}  }

 We study perturbations of exactly tri-bimaximal neutrino mixings under the assumption that they are coming solely from the charged lepton mass matrix. This may be plausible in scenarios where the  mass generation mechanisms of neutrinos and charged leptons/quarks have a different origin. As a working hypothesis, we assume mass textures which may be generated by the Froggatt-Nielsen mechanism for the charged lepton and quark sectors, which generically leads to strong hierarchies, whereas the neutrino sector is exactly tri-bimaximal with a mild (normal) hierarchy.
We find that in this approach, deviations from maximal atmospheric mixing can be introduced without affecting $\theta_{13}$ and $\theta_{12}$, whereas a deviation of $\theta_{13}$ or $\theta_{12}$ from its tri-bimaximal value will inevitably lead to a similar-sized deviation of the other parameter. Therefore, the already very precise knowledge of $\theta_{12}$ points towards small $\stheta \lesssim 0.01$. The magnitude of this deviation can be controlled by the specific form of the charged lepton texture.

\vspace*{.5cm}

\end{titlepage}

\newpage

\renewcommand{\thefootnote}{\arabic{footnote}}
\setcounter{footnote}{0}

\section{Introduction}

Comparing the neutrino masses with the charged lepton and quark masses, they observe a relatively mild hierarchy.  
One can easily see that if one expands the masses and mixing angles of quarks and charged leptons in terms of powers of a single small expansion parameter $\epsilon$.  
In fact, the CKM mixing matrix $V_{\rm CKM}$ \cite{Cabibbo:1963yz,Kobayashi:1973fv} exhibits quark mixing angles of the orders
\begin{equation}\label{equ:CKMangles}
|V_{us}|\sim\epsilon,\quad |V_{cb}|\sim\epsilon^2,\quad
|V_{ub}|\sim\epsilon^3,
\end{equation}
where the quantity $\epsilon$ is of the order of the Cabibbo angle
$\epsilon \simeq \theta_\text{C}\simeq 0.2$. Similarly, for the same value
$\epsilon\simeq \theta_\text{C}$, the mass ratios of the up-type quarks, down-type quarks, and the charged
leptons can be approximated, {\it e.g.}, by
\begin{equation}\label{equ:massratios}
m_u:m_c:m_t \sim \epsilon^6:\epsilon^4:1,\quad
m_d:m_s:m_b \sim \epsilon^4:\epsilon^2:1,\quad
m_e:m_\mu:m_\tau \sim \epsilon^4:\epsilon^2:1,
\end{equation}
where $m_b/m_t\sim\epsilon^2$, $m_\tau/m_b\sim 1$, and $m_t\simeq
175\:\text{GeV}$.  On the other hand, the neutrino mass
spectrum can be roughly written as
\begin{equation}\label{equ:spectra}
m_1:m_2:m_3 \sim \epsilon^2:\epsilon:1,\quad
m_1:m_2:m_3 \sim 1:1:\epsilon,\quad
m_1:m_2:m_3 \sim 1:1:1,
\end{equation}
for the normal hierarchical, inverse hierarchical, and degenerate neutrino mass
spectrum, respectively. While the mixing angles in the quark sector are small,
the lepton  Pontecorvo-Maki-Nakagawa-Sakata (PMNS) mixing matrix $U_{\rm PMNS}$~\cite{Pontecorvo:1957cp} exhibits two large mixing angles and a small (zero) one \cite{Schwetz:2008er}.
It  can be well approximated by the tri-bimaximal (TBM) mixing matrix $U_\text{TBM}$
\cite{Harrison:1999cf} (up to phases) as
\begin{equation}\label{eqU:HPS}
U_\text{PMNS}\approx U_\text{TBM}=
\left(
\begin{matrix}
\sqrt{\frac{2}{3}} & \frac{1}{\sqrt{3}} & 0\\
-\frac{1}{\sqrt{6}} & \frac{1}{\sqrt{3}} & \frac{1}{\sqrt{2}}\\
\frac{1}{\sqrt{6}} & -\frac{1}{\sqrt{3}} & \frac{1}{\sqrt{2}}
\end{matrix}
\right).
\end{equation}
In $U_\text{TBM}$, the solar and the atmospheric angle are given by $\theta_{12}\approx 35^\circ$ and $\theta_{23}=45^\circ$, whereas the reactor angle $\theta_{13}$ vanishes.

The experimental (measured) values can arise as deviations from the TBM ansatz of the neutrino mass matrix \cite{Plentinger:2005kx,Majumdar:2006px,Hochmuth:2007wq,King:2008}, describing nearly TBM lepton mixing \cite{Xing:2002sw}. These perturbations are often motivated by non-Abelian discrete symmetries \cite{Altarelli:2010gt}, 
such as $A_4$ \cite{a4papers}-\cite{Altarelli:2008bg} and $S_4$ \cite{s4papers}.
The main reason is that there exist suitable breakings of such symmetries into subgroups that allow
a neutrino mass matrix exactly diagonalized by TBM. To reconcile with the experimental data, the perturbations can be induced either by
the charged lepton sector or by next-to-leading order contributions to the neutrino mass matrix, or both.
In this work, however, we assume that the mechanism for generating charged fermions and neutrino masses and mixings are different.
This may  not be a remote possibility. In fact, we can assume that Majorana neutrino
masses  are introduced by the lepton number violating Weinberg operator~\cite{Weinberg:1979sa},
\be
\mathcal{O}_W = (\overline{L^{c}} {\rm i} \tau^{2} H)\, (H {\rm i} \tau^2 L) \, ,
\ee
which leads, after electroweak symmetry breaking (EWSB), to Majorana masses
for the neutrinos ($H$ is the SM Higgs doublet). It is well known that this operator implies physics beyond the Standard Model, such as a
heavy neutral fermion leading to the type~I see-saw. Therefore, the origin of the Majorana neutrino mass lies in physics beyond the Standard Model,
including couplings beyond the Standard Model. On the other hand, the other fermion masses can be easily described within the Standard Model, at least as long as
the hierarchies need not to be justified. Or, invoking an $SU(5)$-inspired grand unified framework,  they can be understood
because charged leptons and quarks are arranged in the same GUT multiplets (the neutrinos being singlet of the group).

Therefore, it may be natural to assume that the leading mass generation mechanisms of neutrinos versus charged leptons/quarks are different. 
There is, however, one drawback of this strictly phenomenological separation: since neutrinos and charged leptons come in SU(2) doublets in the Standard Model, 
this strict separation above the EWSB scale might be challenging. For example, left-handed neutrinos and charged leptons will always belong to the same representation of the non-Abelian discrete group.
On the other hand, one or both of the mass generation mechanisms of the neutrinos and charged leptons/quarks may be
implemented at or below the EWSB scale. We will not enter this level of detail, but instead study the phenomenological  consequences of such a model with mass generation mechanisms, which are separated to leading order, can be constructed.

As one example for the charged lepton and quark mass generation, we use the Froggatt-Nielsen (FN) mechanism~\cite{Froggatt:1978nt} in which effective dimension-$n$ mass terms lead to masses proportional to $\epsilon^n$, where $\epsilon$ depends on the flavon vacuum expectation value suppressed by the mass of super-heavy fermions. In this way,
mass matrix textures with $\epsilon$-powers as entries are obtained. Consequently, such a matrix structure contains information on the hierarchy among matrix elements
and goes beyond approaches which use texture zeros. The FN mechanism is a perfectly plausible possibility to generate strong hierarchies. It has, for instance, been used in
\Refs~\cite{Plentinger:2006nb} to construct charged lepton and even neutrino mass textures which can be implemented by discrete flavor symmetries~\cite{Altarelli:2008bg}.
Here we use a similar approach to study the interplay between the charged lepton mass matrix,
generated by this approach, with TBM mixings in the neutrino sector and small deviations from it coming  from the diagonalization
of the charged lepton mass matrix.
Note that in this ansatz the quantity $\epsilon$ determines both the charged lepton and quark mass matrices, which points to a common origin, leading to some form of ``quark-lepton complementarity''. QLC has been studied from many different points of view~\cite{qlcbimax,qlcsumrules,qlcpheno,qlcCabibbo,Chauhan:2006im,qlcRG,qlcmodels}. As in the earlier references~\cite{Plentinger:2006nb}, we implement this quark-lepton complementarity at the Yukawa coupling level, but we
study the implications of random complex order one coefficients, as suggested by the original Froggatt-Nielsen approach, within specific textures leading to interesting deviations from TBM mixings. 

\section{Methods}

We diagonalize the charged lepton and Majorana neutrino mass matrices as
\begin{eqnarray}
M_\ell & = & U_\ell \, M_\ell^{\mathrm{Diag}} \, U_\ell'^\dagger \label{equ:ml} \\
M_\nu & = & U_\nu \, M_\nu^{\mathrm{Diag}} \, U_\nu^T \label{equ:mnu} \, ,
\end{eqnarray}
where $M_\ell^{\mathrm{Diag}}$ and $M_\nu^{\mathrm{Diag}}$ are, up to an overall
mass scale, given by
\equ{massratios} (third relationship) and \equ{spectra}, respectively.
The matrices $U_\ell$, $U_\ell'$, and $U_\nu$ are in general, arbitrary unitary matrices.
By our assumptions, we choose $U_\nu=U_{\mathrm{TBM}}=\hat U(\sin^2\theta_{12}=1/3,\sin^2 \theta_{13}=0,\sin^2 \theta_{23}=1/2)$ with $\hat U$ the mixing matrix in
the standard parameterization~\cite{Nakamura:2010zzi}.
Therefore, the neutrino mass matrix is, together with the choice of the hierarchy in \equ{spectra} and the absolute neutrino mass scale, uniquely determined and assumes special versions of the TBM form.
The lepton mixing matrix is, as usual, given by
\begin{equation}
U_{\mathrm{PMNS}} = U_{\ell}^\dagger U_\nu =  U_{\ell}^\dagger U_{\mathrm{TBM}} \, . \label{equ:upmns}
\end{equation}
Using \equ{upmns} in \equ{ml}, we obtain
\begin{equation}
M_\ell = (U_{\mathrm{TBM}} U_{\mathrm{PMNS}}^\dagger) M_{\ell}^{\mathrm{Diag}} U_{\ell}'^\dagger \, ,
\label{equ:master}
\end{equation}
where $U_{\mathrm{PMNS}}$ is the measured mixing matrix. Note that \equ{master} allows us to construct $M_{\ell}$
for arbitrary $U_{\mathrm{PMNS}}$ deviating from TBM mixings if we fix $U_{\ell}'$.  In the following, we will use  $U_{\ell}'=1$ for the sake of simplicity, which means that $M_\ell=  M_{\ell}^{\mathrm{Diag}}$ if $U_{\mathrm{PMNS}}$ has exactly the TBM form. On the other hand,
any deviation from TBM mixings will lead to a non-trivial mass matrix for $M_{\ell}$ with off-diagonal entries. Our approach therefore corresponds to a particular
form of perturbations of the TBM mixings coming from the charged lepton mass matrix only.

\begin{table}[t]
\begin{center}
\begin{tabular}{lr|rr|rr|rr|rr}
\hline
 & Best-fit & $3 \sigma$ (current) &  $\Delta$ & $3 \sigma$ (2015) & $\Delta$ & $3 \sigma$ (2025) & $\Delta$  & $3 \sigma$ (2035) & $\Delta$ \\
\hline
$\sin^2 \theta_{12}$ & 0.318 & 0.27$\hdots$0.38 & $\epsilon^2$ & \multicolumn{6}{c}{No further experiments planned?} \\
$\sin^2 \theta_{13}$ & 0.013 & $\lesssim$ 0.053 & $\epsilon$ & $\lesssim$ 0.012 & $\epsilon$ & $\lesssim$ 0.001 & $\epsilon^2$ &  $\lesssim 1.5 \cdot 10^{-5}$ & $\epsilon^3$    \\
$\sin^2 \theta_{23}$ &  0.5 & 0.36$\hdots$0.67 & $\epsilon$ & 0.43$\hdots$0.57 & $\epsilon^2$ &  0.47$\hdots$0.53 & $\epsilon^3$ & 0.47$\hdots$0.53 & $\epsilon^3$ \\
\hline
\end{tabular}
\end{center}
\mycaption{\label{tab:dev}\it Current best-fit values for the mixings angles and the $3\sigma$ allowed
ranges~\cite{Schwetz:2008er}, as well as projections for the measurements labeled ``2015''
(mostly Daya Bay and T2K, see \Ref~\cite{Huber:2009cw}), ``2025'' (superbeam upgrades, such as T2HK or LBNE; see \Ref~\cite{Bernabeu:2010rz} for
$\sin^2 \theta_{13}$ and \Ref~\cite{Antusch:2004yx} for $\sin^2\theta_{23}$), and ``2035'' (neutrino factory~\cite{Huber:2003ak,Tang:2009na}); for the projections, $\sin^2 \theta_{13}=0$ is assumed as best-fit value.  
The columns labeled ``$\Delta$'' motivate typical allowed deviations of the angles from their  TBM values in terms of the $\epsilon$ power  such that $\Delta$ is comfortably with the  $3 \sigma$ allowed range for order one coefficients $A$ in \equ{dtbm} (for $\epsilon=0.2$).
}
\end{table}

Since we specify the mass hierarchies in terms of $\epsilon$, we will also motivate the deviations from TBM mixings by powers of $\epsilon$.
We define
\begin{equation}
\Delta_{ij} \equiv \theta_{ij} - \theta_{ij}^{\mathrm{TBM}} = A_{ij} \, \epsilon^{n_{ij}} \, , \quad ij \in \{12, 13,23 \}
\label{equ:dtbm}
\end{equation}
with order one coefficients $\epsilon \le |A_{ij}| \le 1/\epsilon$. We list in \Tab~\ref{tab:dev} the estimated experimental ranges for four different experiment generations, denoted by ``Current'' (current best-fit), ``2015'' (mostly from Daya Bay and T2K), ``2025'' (superbeam upgrades), and ``2035'' (neutrino factory). In this table, we also list plausible deviations in terms of the power of $\epsilon$ from TBM mixing for each angle, motivated by these measurement precisions.
One can clearly read off the table, that each generation of experiments will improve the precision on the mixing angles. If the TBM values are confirmed, $\Delta_{ij}$ will be constrained further, and smaller deviations from TBM mixings will be allowed. Very importantly, $\sin^2 \theta_{12}$ is already very well measured $\propto \epsilon^2$, which has interesting implications -- as we will demonstrate later. We will further on use $\Delta_{ij} = A_{ij} \, \epsilon^{n_{ij}}$ as input assumptions instead of the measurement precisions, but \Tab~\ref{tab:dev} demonstrates that these are closely related.
 Using this hypothesis for $U_{\mathrm{PMNS}}$ in \equ{master}, we can construct $M_\ell$ for each case.
In that case, the mass matrix $M_\ell$  depends on $\epsilon$ (and the absolute mass scale) only. In the spirit of \Ref~\cite{Plentinger:2006nb}, we can then extract a texture from the mass matrix by identifying the leading entry and absorbing the lowest power of $\epsilon$ in the absolute neutrino mass scale. For example, we identify (mass matrix $\rightarrow$ texture):
\begin{equation}
\left(
\begin{array}{ccc}
 \epsilon ^4 & 0 & 0 \\
 0 & \epsilon ^2 & -\frac{\epsilon ^2}{\sqrt{2}} \\
 0 & \frac{\epsilon ^4}{\sqrt{2}} & 1-\frac{\epsilon ^4}{4}
\end{array}
\right)
\rightarrow
\left(
\begin{array}{ccc}
 \epsilon ^4 & 0 & 0 \\
 0 & \epsilon ^2 & \epsilon ^2 \\
 0 & \epsilon ^4 & 1
\end{array}
\right) \, ,
\label{equ:exampletex}
\end{equation}
where ``0'' stands for $\mathcal{O}(\epsilon^5)$.

In the reverse direction, specifying $M_\ell$ as given by the theory and re-diagonalizing it, we should be able, barring ambiguities, to reconstruct the initial hypothesis for the deviation from TBM mixings.\footnote{Note that $U'$ is, by definition, not relevant for this method, because the re-diagonalization will yield $U'=1$ automatically if the mass matrix is constructed that way.}
For example, the texture can then be interpreted in terms of Froggatt-Nielsen-like models with arbitrary (complex) order one coefficients $c_{ij}$:
\begin{equation}
\left(
\begin{array}{ccc}
 \epsilon ^4 & 0 & 0 \\
 0 & \epsilon ^2 & \epsilon ^2 \\
 0 & \epsilon ^4 & 1
\end{array}
\right) \rightsquigarrow \left(
\begin{array}{ccc}
 c_{11} \epsilon ^4 & 0 & 0 \\
  0 & c_{22} \, \epsilon ^2 & c_{23} \, \epsilon ^2 \\
  0 & c_{32} \, \epsilon ^4 & c_{33} \, 1
\end{array}
\right) \, .
\label{equ:fnc}
\end{equation}
In the Froggatt-Nielsen ansatz, the order  of the texture entries is given by a discrete symmetry, whereas
the order one coefficients $c_{ij}$ are random order one entries. Where applicable to test the stability of individual charged lepton textures,
we generate these order one entries $c_{ij} = | c_{ij} | \exp(i \phi_{ij})$ randomly, with $| c_{ij} |$ uniformly distributed between $\epsilon$ and
$1/\epsilon$ and $\phi_{ij}$ in $[0,2 \pi[$.\footnote{In fact, we draw $k$ between $\epsilon$ and $1$ uniformly, and then assign  $| c_{ij} |=k$ or $| c_{ij} |=1/k$ with 50\% probability
each.} Once $M_\ell$, given by \equ{master}, and $M_\nu$, given by \equ{mnu}, are specified, the two mass matrices can be numerically diagonalized,
$U_{\mathrm{PMNS}}$ can be computed, and the mixing parameters can be read off from $U_{\mathrm{PMNS}}$ using re-phasing invariants.
We use the MPT package from \Ref~\cite{Antusch:2005gp} for this part.

\section{Deviations coming from charged lepton mass matrix}

First, we investigate the effect of arbitrary (small) deviations from TBM mixing coming from the charged lepton mass matrix. Our notation ($\Delta_{ij}$) can be easily related to other notations used in the literature. Here we use the parameterization of deviations from TBM from \Ref~\cite{King:2008}, which is in terms of the sines of the angles, not the angles themselves. Therefore, additional pre-factors appear:
\begin{equation}
a \simeq \Delta_{23} \,,\quad s \simeq \sqrt{2} \Delta_{12}\,,\quad r \simeq \sqrt{2} \Delta_{13}~.
\label{equ:arsNotation}
\end{equation}
We use \equ{arsNotation} in \equ{master} and expand the resulting charged lepton mass matrix to second order. In this way, we obtain
\begin{equation}\label{equ:GenMl}
M_\ell^\text{general}\approx
\left(
\begin{array}{ccc}
1-\frac{r^2+s^2}{4} & -\frac{r(1+a)+s(1-a)}{{2}} & \frac{s-r+(r+s) a}{{2}} \\
\frac{r+s}{{2}} & 1-\frac{(r+s)^2/2+2 a^2}{4} & -a-\frac{(r+s)^2/2-s^2}{4} \\
\frac{r-s}{{2}} & a-\frac{(r-s)^2/2-s^2}{4} & 1-\frac{(r-s)^2/2+2 a^2}{4}
\end{array}
\right)
\text{diag}(m_e,m_\mu,m_\tau)~.
\end{equation}
This means that our approach lead to the universal mass matrix \equ{GenMl} which is solely determined by observables. Moreover, the one-to-one correspondence of $M_\ell^\text{general}$ with the experimentally accessible quantities in \equ{arsNotation} in fact allows us to determine the Yukawa couplings in this approach, which can be extracted using \equ{GenMl}.

From the general form of $M_\ell^\text{general}$ in \equ{GenMl}, we can already make some interesting observations. Suppose we wanted to have individual deviations from TBM mixings, such as we only wanted deviations in the 2-3 sector, \ie, $r=s=0$:
\begin{equation}
M_\ell=\left(
\begin{array}{ccc}
 1 & 0 & 0 \\
 0 & 1-\frac{a^2}{2} & -a \\
 0 & a & 1-\frac{a^2}{2}
\end{array}
\right)
\text{diag}(m_e,m_\mu,m_\tau)~.
\label{equ:t23}
\end{equation}
Using \equ{upmns}, it is easy to verify that the first row of $U_{\text{PMNS}}$, which determines $\sin \theta_{13}$ and $\tan \theta_{12}$, remains fixed to
the corresponding TBM entries. This means that we can introduce deviations to $\theta_{23}=\pi/4$ without affecting $\theta_{13}$ and $\theta_{12}$ in this approach. In the reverse direction, the conclusion is that measuring $\theta_{13}=0$ together with a deviation from maximal atmospheric mixing can be trivially interpreted as a perturbation from the charged lepton sector.

On the other hand, if we choose $a=0$, we obtain from \equ{GenMl}:
\begin{equation}\label{equ:Mlc0}
M_\ell^\text{c=0}\approx
\left(
\begin{array}{ccc}
 1-\frac{r^2+s^2}{4} & -\frac{r+s }{{2}} & \frac{s-r }{{2}} \\
 \frac{r+s}{{2}} & 1-\frac{(r+s)^2}{8} & \frac{-r^2-2rs+s^2}{8}  \\
 \frac{r-s}{{2}} & \frac{-r^2+2rs+s^2}{8} & 1-\frac{(r-s)^2}{8}
\end{array}
\right)
\text{diag}(m_e,m_\mu,m_\tau)~.
\end{equation}
Here we see that the matrix is symmetric in deviation of $\theta_{13}$ and $\theta_{12}$. The consequence will be that a perturbation in one of the sectors affects also the other mixing angle, because if it is introduced at the mass matrix level, it cannot be clearly assigned to one of the angles. For example, this can also be seen very specifically in the special case of deviations introduced in the 12-sector of the charged lepton mixing only (CKM-like charged lepton mixings). The sum rule from  \Refs~\cite{King:2005bj} reads, in our notation, $\Delta_{12} \simeq \Delta_{13} \cos \delta$, which shows that the two mixing angles are related.

\section{Mass textures for deviations from TBM mixings}

In this section, we discuss how we can derive charged lepton mass textures in terms of powers of $\epsilon$ for certain perturbations of TBM. We will discuss in the next section, if these textures, if given by a model, indeed lead to the expected deviations from TBM using the Froggatt-Nielsen ansatz as example.
We here parameterize the deviations from TBM mixings in terms of $\epsilon$:
\begin{equation}\label{equ:GenDevE}
\Delta_{13}= R \, \epsilon^{n_{13}}  \, ,
\quad\quad \Delta_{12}= S \, \epsilon^{n_{12}}
\, ,\quad\quad
\Delta_{23}= A \, \epsilon^{n_{23}} \, ,
\end{equation}
where $n_{ij} = 1, 2, 3, \hdots$ is the power of $\epsilon$ which determines the (leading) magnitude of the deviation, and $R$, $S$, $A$ are order one coefficients, \ie,  $\epsilon < |R|, |S|, |A| < 1/\epsilon$. We can also parameterize the charged lepton and neutrino masses in terms of $\epsilon$, see \equ{massratios} and \equ{spectra}, where we use the normal hierarchy case as an example. Applying \equ{GenDevE} and the mass spectra to \equ{GenMl}, we can derive mass textures for individual perturbations of TBM in terms of $\epsilon$-powers; \cf, \equ{exampletex}.

As a first example, let us introduce a deviation from maximal atmospheric mixing assuming $R=S=0$. From \equ{t23}, we read off that for $\Delta_{23}= A \, \epsilon^{n_{23}}$
\begin{equation}
M_\ell=\left(
\begin{array}{ccc}
 \epsilon^4 & 0 & 0 \\
 0 & \epsilon^2 & -A \epsilon^{n_{23}} \\
 0 & A \epsilon^{(2+n_{23})} & 1
\end{array}
\right) \, \overset{n_{23}=1}{\longrightarrow}\left(
\begin{array}{ccc}
 \epsilon^4 & 0 & 0 \\
 0 & \epsilon^2 & \epsilon \\
 0 & \epsilon^3 & 1
\end{array}
\right) \, ,
\label{equ:t23tex}
\end{equation}
where we have computed the texture for $n_{23}=1$ in the last step. As it is obvious from the discussion in  the previous section, the choice of this texture does not significantly affect the TBM values of $\theta_{13}$ and $\theta_{12}$.  We have checked that this conclusion even holds in the presence of arbitrary order one coefficients in the individual texture entries; \cf, \equ{fnc}.
The size of the deviation from TBM is given by the texture entries in \equ{t23tex}, \ie, by choosing a particular texture, one can control the magnitude of the deviation from TBM.

If we want to introduce deviations to the TBM values of $\theta_{13}$ or $\theta_{12}$, we know from \equ{Mlc0}
that these can be not easily generated separately. In general, we find that the texture is determined by the leading (largest) deviation from TBM, or the lowest power $n \equiv \min(n_{13},n_{12}) \ge 1$, as
\begin{equation}
M_\ell \rightarrow \left(
\begin{array}{ccc}
 \epsilon^4 & \epsilon^{2+n} & \epsilon^n \\
 0 & \epsilon^{2} & \epsilon^{2n} \\
 0 & \epsilon^{2+2n} & 1
\end{array}
\label{equ:mastertex}
\right) .
\end{equation}
This leads for $n=1$, $2$, or $3$, to three distinct textures:
\begin{equation}
M_\ell \overset{n=1}{\longrightarrow} \left(
\begin{array}{ccc}
 \epsilon^4 & \epsilon^{3} & \epsilon \\
 0 & \epsilon^{2} & \epsilon^2 \\
 0 & \epsilon^{4} & 1
\end{array}
\right)  \, , \quad
M_\ell \overset{n=2}{\longrightarrow} \left(
\begin{array}{ccc}
 \epsilon^4 & \epsilon^{4} & \epsilon^2 \\
 0 & \epsilon^{2} & \epsilon^{4} \\
 0 & 0 & 1
\end{array}
\right)
\, , \quad
M_\ell \overset{n=3}{\longrightarrow} \left(
\begin{array}{ccc}
 \epsilon^4 & 0 & \epsilon^3 \\
 0 & \epsilon^{2} & 0 \\
 0 & 0 & 1
\end{array}
\right) \,
.
\label{equ:mastertex2}
\end{equation}
It is characteristic for that texture that the smaller the deviation from TBM is, the more so-called texture zeros [$\mathcal{O}(\epsilon^5)$] are in the charged lepton textures. Moreover, the structure as a whole,
represented by powers of $\epsilon$ in the matrix elements, does not change, \ie, the $\epsilon$ power of each matrix element ascends or remains constant while going from case $n=1$ to $n=3$.
For the particular case $n=n_{12}=n_{13}$, the mass matrix reads explicitly to leading order in $\epsilon$
\begin{equation}
M_\ell =
\left(
\begin{array}{ccc}
\epsilon ^4 &\left(-\frac{R}{\sqrt{2}}-\frac{S}{\sqrt{2}}\right) \epsilon ^{2+n} & \left(\frac{S}{\sqrt{2}}-\frac{R}{\sqrt{2}}\right) \epsilon^n  \\
0& \epsilon ^2 & \frac{1}{4} \left(-R^2-2 RS+S^2\right) \epsilon ^{2n} \\
0 & \frac{1}{4} \left(-R^2+2 RS+S^2\right) \epsilon ^{2+2n} & 1
\end{array}
\right)
\label{equ:texttg}
\end{equation}
The case $n_{12} > n_{13}$ can be obtained easily from \equ{texttg} by setting $S=0$, the case $n_{13} > n_{12}$  by setting $R=0$. Here the  ${\cal O}(1)$ coefficients are explicitly taken into account through the $R$ parameter. This mass matrix allows us to interpret future possible measurements of $\theta_{13}$ and $\theta_{12}$ in terms of the three different models $n=1$, $n=2$, or $n=3$, which may be generated by discrete flavor symmetries, \ie, the Yukawa couplings can be measured.

\section{Stability of approach in Froggatt-Nielsen models}

\begin{figure}[tp]
\begin{center}
\includegraphics*[width=\textwidth]{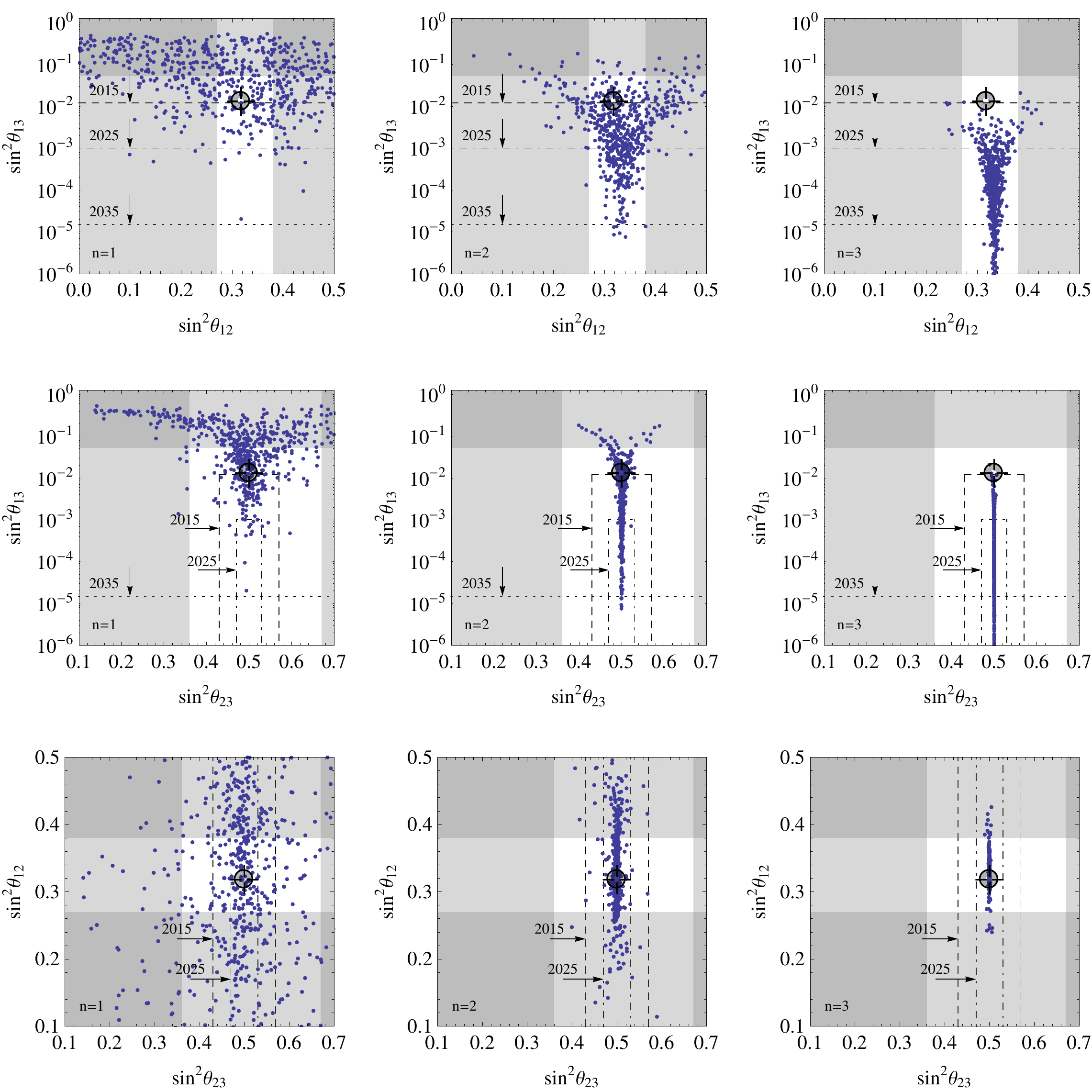}
\caption{\it Correlations among the PMNS mixing angles for 600 unconstrained variation of Yukawa couplings of \equ{mastertex2} for $n=1$ (left column), $n=2$ (middle column), and $n=3$ (right column).
In any panel, the Yukawas are complex parameters with moduli between $1/\epsilon$ and $\epsilon$ and unconstrained phases in $[0,2\pi[$, where $\epsilon=0.2$.
The shaded regions are experimentally excluded and the sensitivity reach of future neutrino experiments on the various mixing angles are marked (the current $3\sigma$ bounds, the expected limit in ``2015'' (dashed line), the limit in ``2025'' (dash-dotted line) and the reach in ``2035'' (dotted line)). The circle indicates the best fit points for the mixing angles,
according to \Tab~\ref{tab:dev}.
}
\label{fig:dev}
\end{center}
\end{figure}

In Froggatt-Nielsen models~\cite{Froggatt:1978nt}, the Yukawa couplings may arise from higher-dimension terms in combination with a flavor symmetry:
\begin{equation}
\mathcal{L}_\mathrm{eff} \sim \langle H \rangle \, \epsilon^n \, \bar{\Psi}_L \Psi_R \, .
\label{equ:fn}
\end{equation}
In this case, $\epsilon$ becomes meaningful in terms of a small
parameter $\epsilon=v/M_F$ which controls the flavor symmetry breaking.\footnote{
Here $v$ are universal VEVs of SM singlet scalar ``flavons'' that break the flavor symmetry,
and $M_F$ refers to the mass of super-heavy fermions, which are charged under the
flavor symmetry. The SM fermions are given by the $\Psi$'s.}
The integer power of $\epsilon$ is solely determined by the
quantum numbers of the fermions under the flavor symmetry. Therefore, the flavor symmetry predicts the mass texture at the Yukawa coupling level, where, in the original FN approach, the coefficients are arbitrarily chosen  complex order one numbers. We therefore test the stability of our textures in this framework using random order one coefficients, as described in \equ{fnc}.

Introducing deviations from $\theta_{23}=\pi/4$ only, \ie, by \equ{t23tex}, does not produce any qualitatively new insight: using this texture, the other two mixing angles are basically not affected. This means that  a deviation $\Delta_{23}$ can indeed be introduced in this model without perturbing the TBM values of $\theta_{13}$ and $\theta_{12}$.

More interesting are the textures in \equ{mastertex2}. We show the effects from constrained variation of Yukawa couplings  for $n=1$ (left column), $n=2$ (middle column), and $n=3$ (right column) in \figu{dev}. Here we can immediately see the implication of \equ{texttg}: since the leading order entry in each mass matrix element dominates, the order one coefficients will induce both $\Delta_{13}$ and $\Delta_{12}$ with the same (leading) order in \equ{GenDevE}, \ie, $n_{13}=n_{12}$.  Therefore, for arbitrarily chosen $\mathcal{O}(1)$ coefficients, a large $\theta_{13}$ comes together with a large deviation of $\theta_{12}$ from its TBM value. In addition, similar-sized deviations of $\theta_{23}$ from maximality are introduced by these charged lepton textures.

In \figu{dev} we show the current bounds and impact of future experiment generations if the TBM values are confirmed. While the precision of $\theta_{13}$ and $\theta_{23}$ do currently not put the case $n=1$ under pressure, the high precision to which $\theta_{12}$ is measured only allows few possibilities for the $n=1$ case. This is illustrated by the first column of \Tab~\ref{tab:dev}, where it is indicated that $\Delta_{12} \propto \epsilon^2$, whereas $\Delta_{13}$ and $\Delta_{23} \propto \epsilon$. Therefore, $\Delta_{12}$ is the discriminator in this case, and $n=1$ can be basically excluded, and with it, large values of $\theta_{13}$, \ie, $\stheta \lesssim 0.01$, as we can read off from the upper middle panel.
 This situation changes for future experiments, where the more precise measurements of $\theta_{13}$ and $\theta_{23}$ will constrain the textures further: the cases ``2015''and ``2025'' enter the parameter space of $n=2$, whereas the Neutrino Factory may even constrain $n=3$ by its extremely well $\stheta$ bound. As one can read off from the right panels, the pressure on $\theta_{13}$ it will then indirectly also constrain the deviations of $\theta_{12}$ and $\theta_{23}$ from their TBM values.

\section{Summary and conclusions}

We have illustrated that introducing deviations to TBM from the charged lepton sector only is a viable and simple approach, which may be motivated by models where the mass generation mechanisms of neutrinos and charged leptons/quarks decouple. Deviations of $\theta_{23}$ can be independently induced by a particular charged lepton texture without affecting the TBM values of $\theta_{13}$ and $\theta_{12}$. The specific form of the texture then controls the magnitude of the deviation from TBM. In a similar way, deviations of $\theta_{13}$ and $\theta_{12}$ can be introduced, where the specific form of the texture again controls the magnitude of the deviations. In this case, however, the deviations of all parameters from TBM are typically introduced with the same magnitude. We have tested and confirmed these conclusions in a Froggatt-Nielsen approach for the charged lepton mass matrix, where we have chosen order one (random) coefficients. In addition, we have shown that the deviations from TBM can be used in particular models to directly measure the Yukawa couplings.

As the main conclusion in this approach, the entanglement between $\theta_{13}$ and $\theta_{12}$ implies that the
currently extremely good measurement of $\theta_{12}$ already exerts pressure on $\theta_{13}$. In fact, plausible textures are obtained for $\stheta \lesssim 0.01$. On the other hand, future measurements of $\theta_{13}$ will limit the deviations of $\theta_{12}$ and $\theta_{23}$ from their TBM values. Deviations of $\theta_{23}=\pi/4$ can, however, be independently introduced, which means that any measured such deviation will not lead to any conclusions for $\theta_{13}$  and $\theta_{12}$ in this approach.

\subsubsection*{Acknowledgments}

This work has been supported by the
Emmy Noether program of Deutsche Forschungsgemeinschaft (DFG), contract no. WI 2639/2-1 [DM, WW].
The work of FP was supported in part by the Italian INFN under the program ``Fisica Astroparticellare''.

\end{document}